\documentclass[aps,amsmath,superscriptaddress,amsfonts,preprintnumbers,amssymb,10pt]{revtex4}
\usepackage{graphicx}
\usepackage{color}
\usepackage{subfigure}
\usepackage{amssymb}
\usepackage{appendix}
\newcommand{\be}{\begin{eqnarray}}
\newcommand{\ee}{\end{eqnarray}}
\usepackage{amssymb,amsfonts,amsthm,indentfirst}
\begin{document}
\title{Insights from intracules and Coulomb holes}
\author{Golam Ali Sekh}
\affiliation{Department of Physics, Kazi Nazrul University, Asansol 713303, India}
\author{Benoy Talukdar}
\email{binoyt123@rediffmail.com}
\affiliation{Department of Physics, Visva-Bharati University, Santiniketan 731235, India}
\author{Supriya Chatterjee}
\affiliation{Department of Physics, Bidhannagar College, EB-2, Sector-1, Salt Lake, Kolkata 700064, India}
\begin{abstract}
We point out that a typical two-electron distribution function in atoms and molecules often called the intracule depends sensitively on the electron-electron repulsion which leads to the so-called Coulomb correlation. The difference between the intracule densities computed by using the correlated and uncorrelated wave functions has been given the name Coulomb hole. Studies in intracule densities  and Coulomb holes form a subject of considerable current interest. We make use of a three-parameter correlated wave function and its uncorrelated limit to study the properties of intracules and Coulomb holes in $He$, $Li^+$, $Be^{2+}$ and $Ne^{8+}$ and thus provide a transparent physical picture for the interplay between the electron-nucleus attraction and electron-electron repulsion in forming Coulomb holes around electrons in atoms of different $Z$ values.
\end{abstract}
\pacs{}
\keywords{}
\maketitle
\section{Introduction}
In classical mechanics one regards waves and particles as completely distinct types of physical entity presumably because waves are continuous and spatially extended, whereas particles are discrete having very little or no spatial extension. Classical physics is a highly successful theory of nature and has led to many astonishing achievements for human beings. For example, using laws of classical mechanics we could build bridges, supersonic jets, develop wireless communication and even put a man on the moon. But if we use classical mechanics to describe the behavior of atoms, molecules and their constituents, we will arrive at wrong results.
\par The subatomic particles sometimes behave as waves \cite{1,2} and also waves sometimes behave as particles \cite{3,4}. This apparently strange behavior is known as the wave-particle duality. In view of this, probabilistic concepts have been used  to develop a new mechanics to describe the properties of subatomic particles. This new mechanics often called the quantum mechanics was formulated by Schr\"{o}dinger \cite{5,6}. In the Schr\"{o}dinger's approach, a function $\psi(\vec{r},t)$, known as the wave function, is assumed to provide a complete description for the behavior of a  particle of mass $m$ moving in the potential $V(\vec{r},t)$. Here $\vec{r}$ stands for the position vector and $t$ represents the time. Thus $\psi(\vec{r},t)$, denotes the quantum mechanical analog of the classical trajectory $\vec{r}(t)$ of the particle. The only interpretative guide to $\psi(\vec{r},t)$ for classical quantum correspondence is that it is large where the particle is likely to be and small elsewhere. This viewpoint was followed by Born \cite{7,8} to provide a statistical interpretation of the wave function by defining  a position probability density
\begin{equation}
P(\vec{r})=\psi(\vec{r},t)\psi^*(\vec{r},t)
\end{equation}
such that $P(\vec{r})dxdydz$ represents the probability of finding the particle in a volume $dxdydz$. Naturally,
\begin{equation}
\int|\psi(\vec{r},t)|^2d^3r=1.
\end{equation}
We note that in close analogy with the demand in (1) and (2), the uncertainty principle, derived by Heisenberg \cite{9,10} is also a statement for the effects of wave-particle duality.
\par The differential equation satisfied by $\psi(\vec{r},t)$ goes by the name Schr\"{o}dinger equation. When the potential in which the particle moves does not involve the time, the solution $\psi(\vec{r},t)$ of the Schr\"{o}dinger equation can be written in the factored form
\begin{equation}
\psi(\vec{r},t)=\phi(\vec{r})e^{-iEt}
\end{equation}
such that the Schr\"{o}dinger equation reduces to \cite{11}
\begin{equation}
[-\frac{1}{2}\bigtriangledown^2+V(\vec{r})]\phi(\vec{r})]=E\phi(\vec{r}).
\end{equation}
In writing (3) and (4) we have used Hartree atomic units in which $\hbar=e=m=1$. Here $\hbar=\frac{h}{2\pi}$, the reduced Planck's constant. We shall use these units throughout. The function $\phi(\vec{r})$ which satisfies the eigenvalue equation (4) is said to represent a stationary state of the particle. The eigenvalue $E$ belonging to the eigen operator $-\frac{1}{2}\bigtriangledown^2+V(\vec{r})$ represents the possible result of precise measurements of the total energy of the particle. In atomic and molecular problems one is interested in the solutions of (4) for some chosen value of $V(\vec{r})$. The existence of a position probability density permits one to calculate the expectation value of the position vector of a particle and we have
\begin{equation}
<\vec{r}>=\int\vec{r}P(\vec{r})d^3r.
\end{equation}
The eigenvalue equation (4) works very satisfactorily for the hydrogen atom and explains the observed spectra fairly accurately. But it was crucial to apply the equation to two-electron atoms and show that the quantum mechanics was not merely an ansatz \cite{12}. It was found that the Schr\"{o}dinger eigenvalue problem is also applicable for the helium atom \cite{13}. The next logical step in this regard was to look for an accurate non-relativistic mathematical description for many-electron atoms. The most common way to do this is to use the Hartree-Fock method \cite{14,15}, which uses a 'mean field' approximation and treats the inter-electronic repulsion in an average way. Here the electronic motion is statistically independent and un-correlated. Such a viewpoint, however, leads to inaccurate results for the ground-state energy of the helium atom and its isoelectronic sequence. Thus studies in the effects of two-electron correlation resulting from electron-electron repulsion have been a subject of interest from the early days of quantum mechanics \cite{16}.

\par In the case of two-electron atoms or ions, the mean value of the inter-electronic separation $r_{12}$ depends very sensitively on the effects of correlation. In particular we have $<r_{12}>_c><r_{12}>_0$ . Here the suffices $c$ and $0$ merely indicate that the expectation values have been calculated using correlated and un-correlated wave functions. Understandably, the correlated  wave functions represents solution of the two- or many-particle Schr\"{o}dinger equation when the potential  includes the inter-electronic repulsion term in addition to the  electron-nucleus attraction. On the other hand, the un-correlated wave function is found by neglecting the inter-electronic repulsion term in the total potential. The many-particle uncorrelated ground-state wave functions are separable in the space coordinates of the particles while the correlated wave functions are entangled. The two-electron probability density for the separable or Hartree-type wave function is given by
\begin{equation}
\rho(\vec{r}_1,\vec{r}_2)=|\psi(\vec{r}_1,\vec{r}_2)|^2=|\psi(\vec{r}_1)|^2|\psi(\vec{r}_2)|^2.
\end{equation}
Admittedly, when we go beyond the Hartree approximation and do not neglect the electronic correlation the two-particle probability density, in addition to $\vec{r}_1$ and $\vec{r}_2$, will depend on the inter-electronic separation $r_{12}=|\vec{r}_1-\vec{r}_2|$.
\par The  distribution function in (6) gives the probability of finding one electron at $\vec{r}_1$ and another at $\vec{r}_2$ and depends sensitively on the correlation effect. Debye \cite{17,18} first realized that for some purposes, the relative position $r_{12}$ of two electrons are more important than their absolute positions. Thus it was necessary to reduce the two-electron density further. The first specific calculation in respect of this was performed by Coulson and Neilson \cite{19} who deduced the expression
\begin{equation}
f(r_{12})=8\pi^2r_{12}[\int_{r_{12}}^{\infty}r_1dr_1\int_{r_1-r_{12}}^{r_1+r_{12}}r_2\psi^2(1,2)dr_2+\int_{0}^{r_{12}}r_1dr_1\int_{r_{12}-r_1}^{r_{12}+r_1}r_2\psi^2(1,2)dr_2]
\end{equation}
for the distribution function of the inter-electronic distance $r_{12}$. In equation (7) $\psi(1,2)$ stands for either the Hartree-Fock  or explicitly correlated wave function. The function $f(r_{12})$ is often called the position-space intracule. Correspondingly, we can also define the momentum-space intracule. In this paper we shall, however, concentrate on the position-space intracule only and introduce the concept of Coulomb hole which is often regarded  as a measure of electron-electron correlation in the charge distribution of atomic and molecular systems. The function $f(r_{12})$ is normalized to unity such that
\begin{equation}
\int_0^{\infty}f(r_{12})dr_{12}=1.
\end{equation}
The normalization (8) is independent of the choice for $\psi(1,2)$. The Coulomb hole has been defined in ref.19 as the difference between $f(r_{12})$ computed by using the correlated wave function and that computed by using the best un-correlated wave function. The intracule and Coulomb hole occupy an important middle ground between the simplicity of electron densities and bewildering complexities of wave functions. In particular, Coulomb hole provides a very simple description of electron-electron interaction in atomic and molecular systems. The object of the present work is to introduce the concepts of intracule and Coulomb hole to graduate students who are familiar with intermediate quantum mechanics only and are yet interested in the problems of contemporary physics. To that end we shall make use of a simple correlated wave function to construct an exact analytic expression for the position-space intracule. We shall then provide some results for the intracules and Coulomb holes of helium and helium-like ions with a view to gain some physical insights for the interplay between the electron-nucleus and electron-electron interactions in forming the Coulomb hole.
\par In the next section we introduce a simple three- parameter correlated wave function and examine its uncorrelated limit.  We use these wave functions to construct exact analytic expressions for the position-space correlated and uncorrelated intracules. The wave function of our interest was  parametrized by Mukoyama \cite{20} for helium and helium-like ions up to $Z=82$ and written in terms of Hyllerasas variables \cite{16} $s=r_1+r_2$, $t=r_1-r_2$  and $u=r_{12}$ for which $-u\leq t\leq u \leq s\leq \infty$. We devote section 3 to calculate results for intracules and Coulomb holes in $He$, $Li^+$, $Be^{2+}$ and $Ne^{8+}$ and study in some detail how the interplay between the electron-nucleus and electron-electron interactions affects the position and size of such holes which are responsible for inter-electronic repulsion. Finally, in section 4 we summarize our outlook on the present work and make some concluding remarks.
\section{THE POSITION-SPACE INTRACULES}
In terms of the Hylleraas variables or coordinates the expression for the position-space intrcule in (7) can be written as \cite{19}
\begin{equation}
f(u)=\pi^2u\int_u^{\infty}ds\int_{-u}^{u}dt(s^2-t^2)\psi^2(1,2).
\end{equation}
The three-parameter electronic wave function of helium-like atoms as given in ref. 20 is given by
\begin{equation}
\psi(s,t,u)=Ne^{-\frac{ks}{2}}(c_0+c_1 k u+c_2 k^2 t^2)
\end{equation}
with
\begin{subequations}
\begin{eqnarray}
N=\frac{Z^3}{\pi}(1-1.069372/Z+0.2096187/Z^2+0.01328097/Z^3+0.04569446/Z^4),
\end{eqnarray}
\begin{eqnarray}
k = 2Z(1-0.1598726/Z-0.02149917/Z^2-0.05152167/Z^3-0.001302860/Z^4),
\end{eqnarray}
\begin{eqnarray}
c_0=1,
\end{eqnarray}
\begin{eqnarray}
c_1=0.1221154/Z+0.06801699/Z^2+0.02360930/Z^3-0.005988795/Z^4
\end{eqnarray}
\mbox{and}
\begin{eqnarray}
c_2=0.009255006/Z+0.01333793/Z^2+0.006919765/Z^3+0.01799133/Z^4.
\end{eqnarray}
\end{subequations}
The energy expectation values computed by using the wave function in (10) were shown to be in satisfactory agreement with  results obtained from more elaborate calculation \cite{21}.
\par From (9) and (10) it is straightforward to obtain a closed form expression
\begin{equation}
f_c(u)=\frac{N^2}{k^3}u^2e^{-ku}(4+\sum_{i=1}^6a_iu^i)
\end{equation}
for the intracule density with
\begin{subequations}
\begin{eqnarray}
a_1=8c_1+4k,
\end{eqnarray}
\begin{eqnarray}
a_2=4c_1^2+\frac{8c_2}{3}+8kc_1+\frac{4k^2}{3},
\end{eqnarray}
\begin{eqnarray}
a_3=\frac{8c_1c_2}{3}+4c_1^2k+\frac{8c_2k}{3}+\frac{8c_1k^2}{3},
\end{eqnarray}
\begin{eqnarray}
a_4=\frac{4c_2^2}{5}+\frac{8c_1c_2k}{3}+\frac{4 c_1^2 k^2}{3}+\frac{8 c_2 k^2}{15}
\end{eqnarray}
\begin{eqnarray}
a_5=\frac{4c_2^2k}{5}+\frac{8c_1c_2k^2}{15},
\end{eqnarray}
\mbox{and}
\begin{eqnarray}
a_6=\frac{4c_2^2k^2}{35}.
\end{eqnarray}
\end{subequations}
In writing (12) we used $f_c(u)$ for $f(u)$ merely to indicate that the intracule density has been calculated by using a correlated wave function. Such a density obtained from the corresponding un-correlated wave function will be denoted by $f_0(u)$.
\par If we approximate $N$ and $k$ in (11a) and (11b) by their leading terms and use $c_1=c_2=0$, then the correlated wave function (10) gives the well known separable form of the un-correlated  two-electron wave function
\begin{equation}
\psi(\vec{r}_1,\vec{r}_2)=\frac{Z^3}{\pi}e^{-Z(r_1+r_2)},
\end{equation}
which for the choice $Z_{eff}=1.69$ for $Z$ yields the ground state energy of helium atom as -2.85 atomic units. This result is about $1.9\%$ higher than the experimental value for the minimum energy we need to remove both electrons from the helium atom \cite{11}. However, the expression
\begin{equation}
f_0(u)=Z^3u^2e^{-2Zu}(\frac{1}{2}+Zu+\frac{2Z^2u^2}{3})
\end{equation}
computed by using the un-correlated wave function also gives reasonable numerical results \cite{22} for $Z_{eff}=1.69$. We shall now make use of (12) and its un-correlated  limit (15) to study the behavior of $f_c(u)$, $f_0(u)$ and the Coulomb hole defined by
\begin{equation}
\Delta f(u)=f_c(u)-f_0(u)
\label{eq16}
\end{equation}
as a function of $u$.

Many authors have focused on the short-range behavior of intracule densities. Relatively recently,  Cioslowski et al. \cite{23} performed a detailed analysis in respect of this. However, it appears that Thakkar and Smith \cite{24} first demonstrated that electron-electron cusp condition \cite{25} as written in terms of two-electron wave function and its first derivative at the coalescence point implies a corresponding condition for the spherical average of the intracule density $h(u)$  given by
\begin{eqnarray}
\lim_{u \to 0}\left[\frac{1}{2} h'(u)/ h(u) \right]=\frac{1}{2}. 
\label{eq17}
\end{eqnarray}
For the results in (12) and (15) we can write, in general, $h(u)=f(u)/u^2$  . From (15) it is easy to verify that, for the uncorrelated intracule density, the limit in (17)  is equal to zero.  For the correlated result in (12) we obtained a non-zero value of this limit given by
\begin{eqnarray}
\lim_{u \to 0}\left[\frac{1}{2} h'_c(u)/ h_c(u) \right]&=&\frac{1}{Z^7} \left(
0.244231 Z^7+0.096988 Z^6+0.020219 Z^5\right.\nonumber\\&-&\left. 0.035034 Z^4-0.006427 Z^3-0.002352 Z^2+0.000556 Z+0.000016 \right). 
\label{eq18}
\end{eqnarray}
This observation confirms that the presence of $u^{-1}$ singularity in the Hamiltonian of an electronic system is the physical origin of the electron-electron cusp \cite{26}.

\section{NUMERICAL RESULTS FOR INTRACULES AND COLOUMB HOLES}
It is straightforward to use (12), (15) and (16) to study the properties of intracule densities and Coulomb holes for atoms  in the helium isoelectronic sequence. We have pointed out that the choice $Z_{eff}=1.69$ for helium gives fairly good results for $f_0(u)$ for the helium atom. Thus to compute the corresponding results for $f_c(u)$ we can use the same value of the effective nuclear charge. The question now arises - how should we choose the value of $Z_{eff}$ to compute intracule densities for high-Z helium-like ions? In this context we note that for the K-shell electrons of helium the choice of the atomic shielding or screening constant given by $\sigma=Z-Z_{eff}=0.31$ is not far off from the reality. All helium-like ions have electrons in the  K-shell only. Slater \cite{27} has given a simple rule for the choice of screening constants for approximate wave function for all atoms in all stages of ionization. According to Slater's rule the screening constant of the K-shell electrons is given by $\sigma=0.3$. In view of this we shall use this value of $\sigma$ to determine the effective nuclear charge for all atoms in the helium isoelectronic sequence. While computing the values of $f_0(u)$ and $f_c(u)$ it should be noted that (15) always satisfies (8) but $f_c(u)$ in (12) will satisfy the normalization condition only when the correlated wave function is exact \cite{22}. Since in this work we have used only an approximate wave function, we need to scale $f_c(u)$ appropriately in order to satisfy (8).
\par In figure 1a we display the results of $f_0(u)$ and $f_c(u)$ for helium as a function of $u$. The corresponding results for the Coulomb hole are shown in figure 1b.
\begin{figure}[!ht]
\begin{center}
\subfigure[]{\includegraphics[width=8cm, height=6cm]{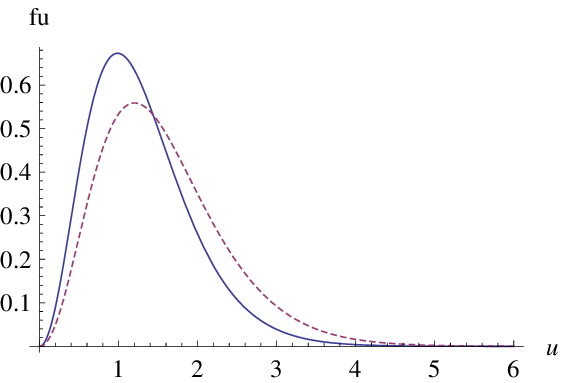}}
\hspace*{0.5cm}
\subfigure[]{\includegraphics[width=8cm, height=6cm]{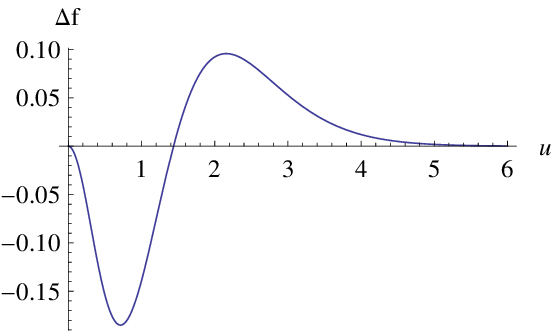}}
\caption{(a) Intracule densities $f(u)$ of $He$ as a function of $u$. (b) Coulomb hole corresponding to the intracule densities in Figure 1a.}
\end{center}
\end{figure}
The solid and dashed lines in figure 1a denote variation of $f_0(u)$ and $f_c(u)$ as a function of $u$. The maximum of the dashed curve for the intracule density $f_c(u)$ calculated by using correlated wave function \cite{20} is shifted towards the right when compared with the maximum of the solid curve giving the variation of $f_0(u)$. In addition the maximum of the dashed curve falls below that of the solid curve. The coordinates $(u,f(u))$ of the maxima of the solid and dashed curves are given by $(0.99, 0.6702)$ and $(1.2, 0.5677)$ respectively. For $u<1.749$ the numbers for $f_c(u)$ fall below those of $f_0(u)$ while for $u>1.749$ the dashed curve lies above the solid one. The results presented in this figure are in good agreement with the corresponding values computed by Curl and Coulson \cite{22} using 10-term Hylleraas wave function \cite{16}. In this respect the works of Benesch \cite{28} and of Koga et al. \cite{29} deserve special attention.

\par Like the two-particle and the intracule densities, the single-particle density is also affected by the electron-electron repulsion \cite{30}. But the response of intrcule density to electronic correlation appears to be more pronounced compared to that of the single-particle charge density. The single particle charge density $\rho(\vec{r_1})$ is obtained from $\rho(\vec{r}_1,\vec{r}_2)$ by integrating over the variable $\vec{r}_2$. Some physical information about the system might be lost during the process of integration such that $\rho(\vec{r}_1)$ is less susceptible to correlation effect when compared with the response intracule density to inter-electronic repulsion \cite{31}. Figure 1b shows the curve for the Coulomb hole as a function of $u$ corresponding to the intracule densities of figure 1a. The shape of our curve is exactly the same as that found by Cioslowski and Liu \cite{32} in their exact calculation. For example, it begins from zero, takes up negative values to have a minimum and then moves upward, reaches a maximum (positive), and tends towards zero. But our results for the minimum and maximum of $\Delta f$ are only in qualitative agreement with those of ref.32.
\begin{figure}[!ht]
\begin{center}
\subfigure[]{\includegraphics[width=8cm, height=6cm]{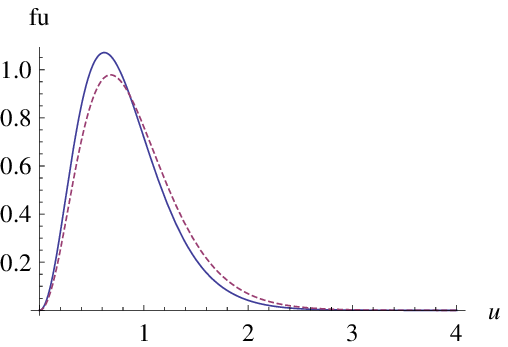}}
\hspace*{0.5cm}
\subfigure[]{\includegraphics[width=8cm, height=6cm]{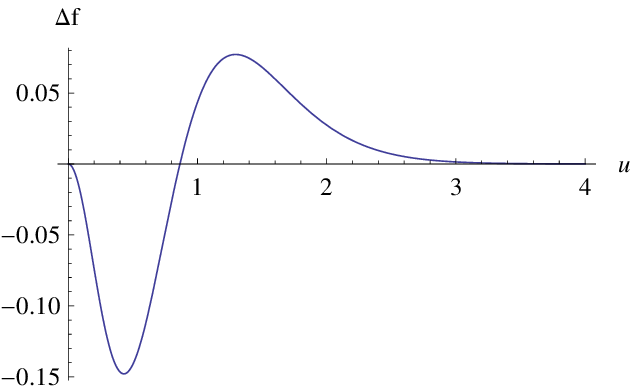}}
\caption{(a) Intracule densities $f(u)$ of $Li^+$ as a function of $u$. (b) Coulomb hole corresponding to the intracule densities in Figure 2a.}
\end{center}
\end{figure}
\begin{figure}[!ht]
\begin{center}
\subfigure[]{\includegraphics[width=8cm, height=6cm]{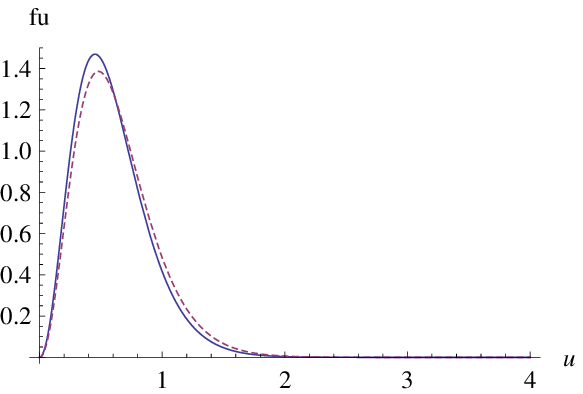}}
\hspace*{0.5cm}
\subfigure[]{\includegraphics[width=8cm, height=6cm]{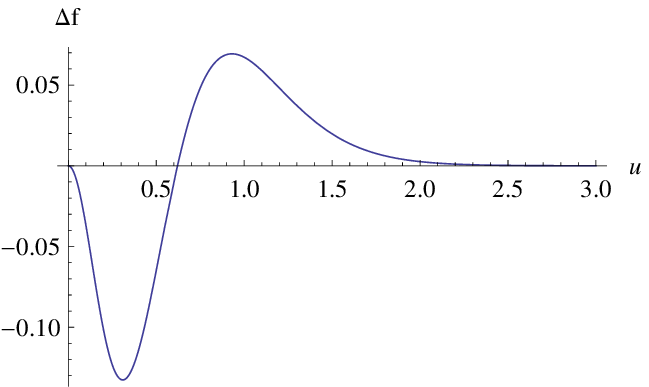}}
\caption{(a) Intracule densities $f(u)$ of $Be^{2+}$ as a function of $u$. (b) Coulomb hole corresponding to the intracule densities in Figure 3a.}
\end{center}
\end{figure}
It will, however, be highly instructive to use our pedagogic approach to examine how do the behavior of position-space intracules and that of Coulomb holes change as the nuclear charge increases. To that end we present in figures 2 and 3 the plots of intracules and Coulomb holes for the ions $Li^+$ and $Be^{2+}$. In particular, the curves in figures 2a an 3a give the variation of position-space intracules with $u$ whereas those in figures 2b and 3b denote similar variation for the corresponding Coulomb holes. Looking closely into the curves for intracules in figures 1a -3a we see intracule densities become more and more peaked as we move from helium to $Be^{2+}$ as well as the peaks move towards the left. Understandably, this is a kind of squeezing of the distribution function and has its dynamical origin in the increase of effective nuclear charge of atomic systems as we go along the periodic table. The observed squeezing, whatsoever, is independent of the correlation effect. As we go from $He$ to $Be^{2+}$ the deviation between the solid and dashed curves gradually diminishes. This implies that as the relative magnitude of the atomic electron-nucleus attraction increases, the effect of electron-electron correlation on the two-electron distribution function gradually diminishes. The positions, depths, heights, minima and maxima of the curves  in figures 1b-3b for Coulomb holes are given in table 1.
\begin{table}[htb!]
\begin{center}
\begin{tabular}{|l|l|l|}
\hline
Atom/Ion & Minima & Maxima \\
\hline
 & Position\;\;Depth & Position\;\;Height\\
\hline
$He$ & 0.7680\;\;\;\;0.1784 & 2.1500\;\;\;\;0.0958\\
\hline
$Li^+$ & 0.4368\;\;\;\;0.1477 & 1.3060\;\;\;\;0.0778\\
\hline
$Be^{2+}$ & 0.3168\;\;\;\;0.1324 & 0.9216\;\;\;\;0.0694\\
\hline
\end{tabular}
\caption{The position, depth and height of the intracules.}
\end{center}
\end{table}
\begin{figure}[!ht]
\begin{center}
\subfigure[]{\includegraphics[width=8cm, height=6cm]{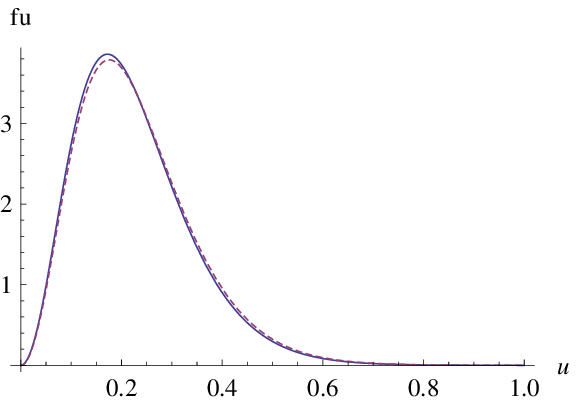}}
\hspace*{0.5cm}
\subfigure[]{\includegraphics[width=8cm, height=6cm]{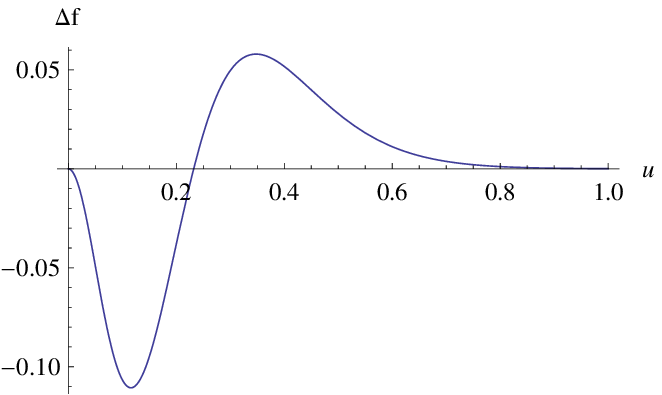}}
\caption{(a) Intracule densities $f(u)$ of $Ne^{8+}$ as a function of $u$. (b) Coulomb hole corresponding to the intracule densities in Figure 4a.}
\end{center}
\end{figure}
The results presented in table 1 indicate that, in close analogy with the behavior of intracule densities, minima and maxima of the Coulomb holes shift to the left as we move to higher $Z$ values. But their depths and heights decrease with $Z$ leading gradually to relatively weak Coulomb holes.  One may be interested to know, in the  helium iso-electronic sequence, what is the highest value of $Z$ after which the Coulomb hole becomes vanishingly small. For a plausible answer to this query we present in figures 4a and 4b the results for intracule densities and Coulomb hole of $Ne^{8+}$ as a function of $u$. The solid and dashed curves in figure 4a are hardly discernible. Consequently, the corresponding Coulomb hole in figure 4b is very much contracted and extremely weak. Thus we can infer that correlation effect is really insignificant in helium-like atoms after $Z=10$.
\begin{figure}[!ht]
\begin{center}
\subfigure[]{\includegraphics[width=8cm, height=6cm]{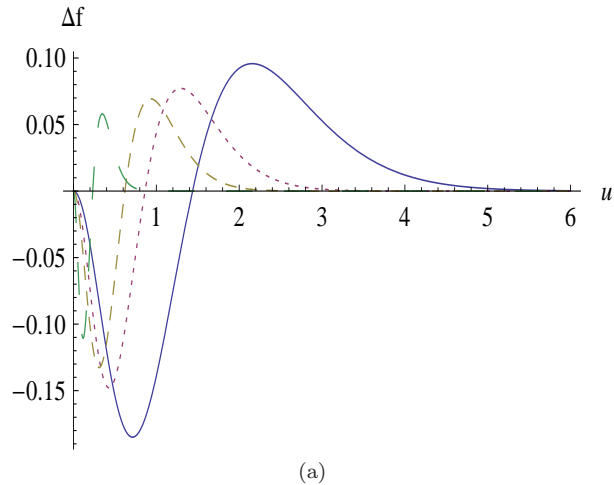}}
\caption{Contraction of Coulomb holes for Z=2 (blue), 3 (pink), 4 (brown) and 10 (green).}
\end{center}
\end{figure}
\par To have a quick look into the relative magnitude of the Coulomb holes in the atomic systems considered by us we display  the curves of figures 1b - 4b in a single plot (figure 5). Clearly, the size of the Coulomb hole for helium represented by the solid (blue) curve is much bigger than that of any other hole presented in the figure. The curves with small (pink), medium (brown) and large (green) dashings portray the Coulomb holes in $Li^+$, $Be^{2+}$ and $Ne^{8+}$ respectively. From these curves it is apparent that, in addition to being contracted towards the nucleus, the size of each Coulomb hole decreases gradually as we go to high-$Z$ atoms.
\par From (18) we have found the values of the electron-electron cusps for $He$, $Li^+$, $Be^{2+}$ and $Ne^{8+}$ as 0.3005, 0.2811, 0.2713 and 0.2541 respectively. These numbers are somewhat inferior to the corresponding results found by Thakkar and Smith \cite{24} by using a 20-parameter Hyllaraas type wave function.

\section{CONCLUDING REMARKS}
The term intracule currently used for two-electron distribution function with $r_{12}$ = constant seems to make its debut in Eddington's text book \cite{18} but it was Coleman \cite{33} who popularized the term in modern electronic structure theory. The correlation analysis based on the intracule density functional developed initially for helium and helium-like atoms has now been extended to molecular systems \cite{34,35}. In this paper we have, however, made use of a simple analytical model to present results for position-space intracules and Coulomb holes for a few atoms in the helium iso-electronic sequence and thereby tried to gain some physical weight for their response to electron-nucleus attraction.

\par The position-space intracule provides a distribution function for $\vec{r}_{12}$. In the same way, the momentum-space intracule $I(\vec{\nu})$ represents the probability distribution function of a relative momentum vector $\vec{\nu}=\vec{p}_{1}-\vec{p}_{2}$ \cite{36}. We note that studies in the momentum-space have been envisaged following similar methods as used in dealing with position-space intracules \cite{37}. The position-space intrcule is a function of inter-particle separation only and as such it cannot provide any information regarding  distribution of the center of mass of the pair. One way we can extract such information is through the extracule density, $E(\vec{r})$, which describes the probability of finding the center of mass of a pair \cite{38}. In the recent past Proud and Pearson \cite{39} proposed a distribution function which involves both intracule and extracule coordinates simultaneously. As opposed to the analytic approach followed in this work, computation of pair densities for high $Z$ atoms calls for the use of purely numerical routines \cite{40,41}.
\vskip0.5cm
{\bf{Acknowledgement:}}
One of the authors (Golam Ali Sekh) would like to thank Department of Science and Technology, Govt. of India for a research grant (CRG/2019/000737).

\end{document}